\documentclass[doublecol]{epl2} 
\usepackage{graphicx} 
\usepackage{dcolumn} 
\usepackage{bm} 
\usepackage{float} 
\usepackage{bbm} 
\usepackage{color}
\usepackage{amsfonts}
\usepackage{lmodern} 
\usepackage{amsmath,amssymb,mathtools}

\usepackage[colorlinks=true,citecolor=blue]{hyperref}
\hypersetup{colorlinks=true,citecolor=blue,linkcolor=red,urlcolor=blue}

\newcommand\beq{\begin{equation}}      
\newcommand\beqnn{\begin{eqnarray*}}   
\newcommand\beqa{\begin{eqnarray}}     
\newcommand\beqann{\begin{eqnarray*}}  

\newcommand\eeq{\end{equation}}        
\newcommand\eeqnn{\end{eqnarray*}}     
\newcommand\eeqa{\end{eqnarray}}       
\newcommand\eeqann{\end{eqnarray*}}    

\def\etal{\emph{et al. }}
\newcommand\bi{\begin{itemize}}
\newcommand\ei{\end{itemize}}

\def\etal{\emph{et al. }}

\newcommand{\eref}[1]{(\ref{#1})}

\newcommand{\Eref}[1]{Eq.~(\ref{#1})}

\newcommand{\al}[1]{\begin{align} #1 \end{align}}
\def\l0{L}

\def\L0{\bm L_0}

\newcommand*\colvec[3][]{
    \begin{pmatrix}\ifx\relax#1\relax\else#1\\\fi#2\\#3\end{pmatrix}
}

\definecolor{purple}{rgb}{0.7, 0., 0.8}

\title{Interplay of quenching temperature and drift in Brownian dynamics}

\author{H. Khalilian\inst{1} \and 
M.R. Nejad\inst{2}
\footnote{Email: mehrana.raeisiannejad@physics.ox.ac.uk}
\and A.G. Moghaddam\inst{3}  \footnote{Email: agorbanz@iasbs.ac.ir}
\and C.M. Rohwer\inst{4,5}}
\shortauthor{H. Khalilian \etal}

\institute{                      
  \inst{1}  School of Nano Sciences, Institute for Research in Fundamental Sciences (IPM), P. O. Box 19395-5531, Tehran, Iran\\
  \inst{2} The Rudolf Peierls Centre for Theoretical Physics, 1 Keble Road, Oxford, OX1 3NP, UK \\ 
    \inst{3} Department of Physics, Institute for Advanced Studies in Basic Sciences (IASBS), Zanjan 45137-66731, Iran\\   
  \inst{4} Max Planck Institute for Intelligent Systems, Heisenbergstr. 3, 70569 Stuttgart, Germany \\  
  \inst{5} 4th Institute for Theoretical Physics, University of Stuttgart, Pfaffenwaldring 57, 70569 Stuttgart, Germany
  }

\pacs{05.40.-a}{Fluctuation phenomena, random processes, noise, and Brownian motion}
\pacs{05.70.Ln}{Non-equilibrium and irreversible thermodynamics}

\abstract{
We investigate the non-equilibrium evolution of ideal Brownian particles confined between two walls, following simultaneous quenches of the temperature and a constant external force. 
We compute (analytically and in numeric simulations) the post-quench dynamics of the density and the pressure exerted by the particles on the two walls perpendicular to the drift force. For identical walls, symmetry breaking associated with the drift gives rise to unequal particle densities and pressures on the two walls. While the pressure on one wall increases monotonically after the quench, on the other wall, depletion causes a non-monotonic dynamics with an overshooting at finite times, before the long-term steady-state value is reached.
For walls immersed in a Brownian gas, the effective interaction force changes sign from repulsive at short times to attractive at late times. These findings have potential applications in various soft matter systems or fluids with charged Brownian particles, as well as carrier dynamics in semiconducting structures.
}

\begin{document}

\maketitle


\section{Introduction} 
Non-equilibrium processes are at the heart of various areas physical phenomena, and have particular relevance to condensed matter physics \cite{zia}.
Mathematical descriptions of such processes fall broadly into two main categories: models of non-equilibrium steady states involving continuous energy pumping, e.g. for active matter, and models focusing on the explicit time evolution due to external driving fields or \emph{quenches} (sudden changes) of some systemic parameters. In quench problems, the central questions concern transient behavior of the system as it approaches a steady state, as well as the nature of this  state. 
While quench-induced dynamics in closed quantum mechanical systems have attracted much interest in recent years \cite{polkovnikov,calabrese}, studies of post-quench dynamics have their roots in the statistical physics of classical systems, and particularly in critical phenomena \cite{hohenberg}.

The presence of globally conserved quantities imposes constraints on the time evolution of non-equilibrium systems. Within coarse-grained models \cite{hohenberg,kardar-book,onuki-book}, such non-trivial dynamics give rise to novel non-equilibrium phenomena such as spinodal decomposition following temperature quenches towards a symmetry-broken phase \cite{binder,furukawa,bray}. Indeed, driven systems with conserved quantities such as the particle number, exhibit long-ranged correlations \cite{spohn1983,dorfmankirkpatricksengers1994,mukamelkafri1998}, which can in turn give rise to fluctuation-induced forces (FIFs) \cite{casimir,kardar99} in non-equilibrium systems \cite{fifq,podgornik16,tfif,aminovkardarkafri2015,rao18}. Sudden quenches, e.g., of the temperature, can modify or generate correlations and FIFs in fluid media (see, e.g., Refs.~\cite{gambassi2008EPJB,deangopinathan2010PRE,rohwer17,gross2018surface,rohwer19,gross2019ModelBcrit}). Remarkably, transient long-ranged correlations even emerge following quenches in fluids that have uncorrelated (i.e., force-free) steady states \cite{rohwer17,rohwer18,rohwer19}. However, temperature quenches additionally modify the density field between objects immersed in a fluid. Corresponding ``density-induced'' non-equilibrium forces, which have been predicted and observed in conserved fluid media \cite{rohwer18}, exist even in non-interacting fluids, and are longer-ranged than their fluctuation-induced counterparts, which vanish in non-interacting systems. 

In this letter, we consider such a system of non-interacting Brownian particles with conserved density, subjected to simultaneous quenches of temperature and of a constant external drift force applied to the system. At large scales, this problem can be described in terms of the post-quench density, which, in the presence of an external force, obeys the diffusion-drift equation. The quench thus involves sudden changes of both the diffusion coefficient (which depends linearly on temperature) and the drift velocity. We expect such a description to be relevant for diffusive media in which quenches of the temperature (or effective temperature~\cite{loiEffectiveT2008,bizonne}) can be realized. Examples include (thermal) Brownian fluids~\cite{karatzas1998brownian}, but also active Brownian fluids~\cite{solon_pressure_2015PRL,berthier2015epl}, e.g., comprised of driven colloids which self-propel due to illumination \cite{buttinoni2012active}, dilute vibrated granular matter~\cite{kudrolli2004size,keys2007measurement}, or dilute colloidal systems in gravitational fields~\cite{senis2001systematic,bechinger2014gravitaxis}.
Intriguingly, similar physics emerges in semiconducting structures in which, despite the quantum nature of charge carriers, electron/hole transport obeys classical diffusion-drift equations \cite{neamen2012-book,hansch2012-book,ancona1989quantum,degond2005quantum,kleinert2010quantum}.

We study the time evolution of non-equilibrium forces on parallel walls following the quenches, analytically and with numerical simulations. The interplay of the temperature quench and drift gives rise to complex and rich dynamics. The pressure differs on the two walls, which are positioned perpendicular to the drift direction. Further, the pressure on one of the walls evolves non-monotonically in time, reaching a maximum before decaying to the steady state. Then, by considering Brownian particles both inside and outside the walls, we show that the effective interaction force between the walls changes sign from repulsive at short times to attractive at late times times following the quench. 

\section{System and simulation model}
\begin{figure}[t!]
\centering
\includegraphics[width=0.70\linewidth]{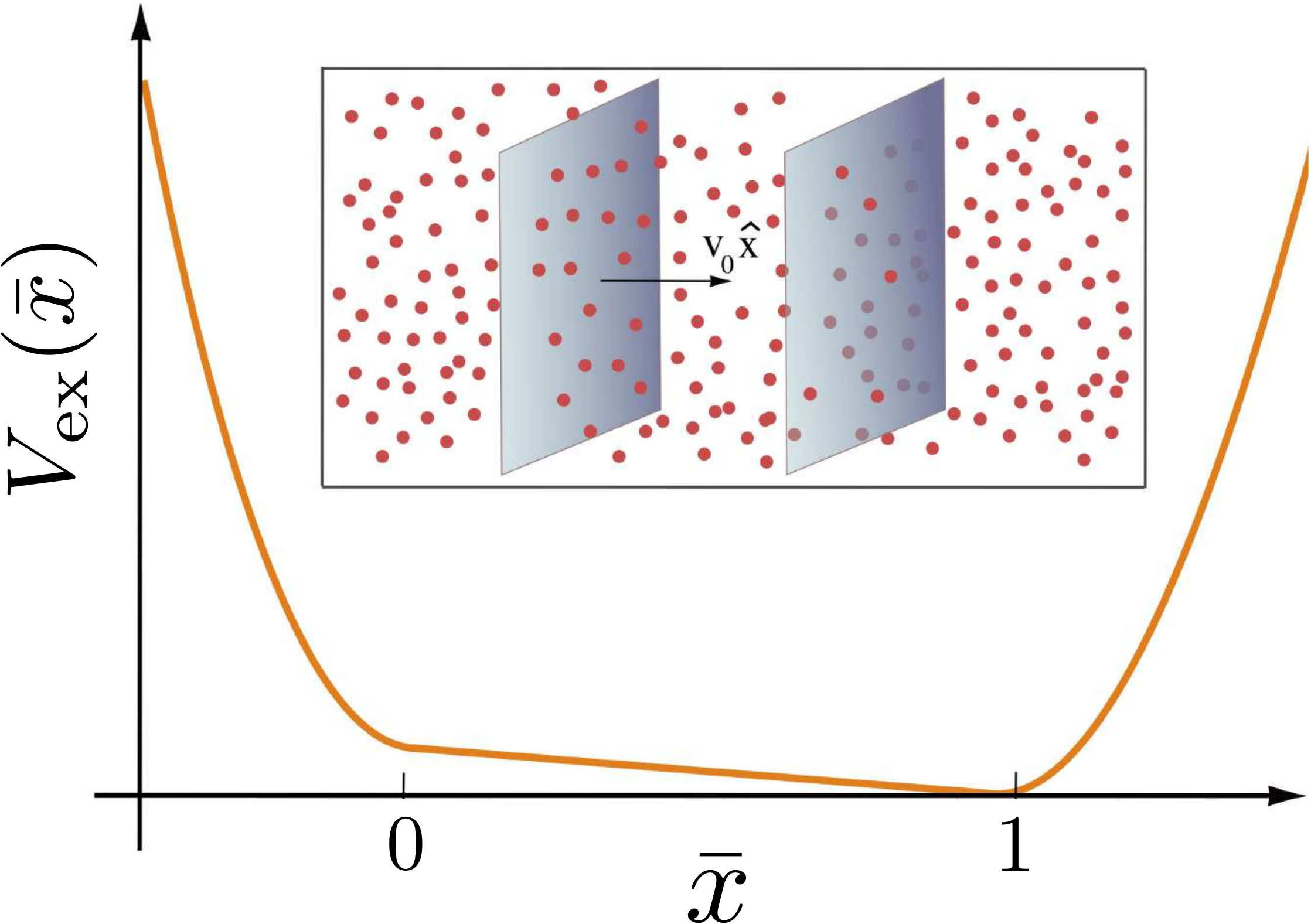}
\caption{(Color online) External potential $V_{\rm ex}({\bar x})$, consisting of a linear term (generating a drift velocity $v_0$ between the two walls),  and the harmonic wall potentials. 
Inset sketch: two parallel plates at $\bar x=x/L=0,1$ immersed in a Brownian gas. }
\label{fig1}
\end{figure}

We consider a system of non-interacting overdamped Brownian particles confined in the $x$ direction between two walls at $x=0,L$ (see Fig.~\ref{fig1}). While our system lives in $d$ dimensions, translational invariance along the walls implies that dynamics is effectively 1D. Each particle $i = 1,\ldots,N$ is governed by a (de-dimensionalized) Langevin equation,
\al{
\label{one}
\frac{d \bar{x}_i}{d \bar{t}}&= -\frac{d \bar{V}_{\rm ex}}{d \bar{x}}|_{\bar{x}_i}+\eta_i(\bar{t}),\\ 
\bar{x}&=x/L,\:\:\:\:\bar{V}_{\rm ex}=V_{\rm ex}/(k_B T),\:\:\:\: \bar{t}=D t/L^2.\nonumber
}
The wall separation $L$, thermal energy $k_BT$, and collective diffusion coefficient $D$ have been used to define dimensionless time $\bar{t}$, positions $\bar{x}_i$, and external potential $\bar{V}_{\rm ex}$. The random force $\eta_i$ is a Gaussian white noise obeying $\langle\eta_i(\bar{t}) \eta_j(\bar{t}^{\prime})\rangle= \delta(\bar{t}-\bar{t}^{\prime}) \: \delta_{ij}$
where $i$ and $j$ are particle indices. 

At time $\bar{t}=0$ a linear external potential is turned on (quenched) between the walls, giving rise to drift-like forces on all particles for $\bar{t}>0$. 
Assuming a harmonic form for the wall potential, $V_{\rm wall}(\bar{x})=(\lambda L^2/2)\, \Theta(-\bar{x})\bar{x}^2$,
the total external potential reads
\begin{eqnarray}
V_{\rm ex}(\bar{x}) = \begin{cases}
 V_{\rm wall}(\bar{x})+ v_0 L/\mu,&\bar{x}<0,\\
  (v_0 L /\mu)\, (1-\bar{x}),&0<\bar{x}<1,\\
 V_{\rm wall}(1-\bar{x}),&\bar{x}>1,
\end{cases}
\label{external}
\end{eqnarray}
as illustrated in Fig.~\ref{fig1}. 
Here, $v_0$ indicates the drift velocity of the particles due to the quench of the external force, and $\mu=D/k_BT$ is the mobility of the particles at thermal conditions with temperature $T$. 
The Langevin equation \eref{one} thus becomes
\begin{equation}
\frac{d \bar{x}_i}{d \bar{t}}= 2\bar{v}_d -\bar{\lambda} \left[  \bar{x}_i   \Theta(-\bar{x}_i)-  (\bar{x}_i-1) \Theta(\bar{x}_i-1)\right ]
+ \eta_i (\bar{t}),
\label{onel}
\end{equation}
where $\bar{v}_d=v_0L/(2\mu k_BT)$ and $\bar{\lambda}=\lambda L^2/k_BT$ are dimensionless parameters representing the strengths of the drift- and wall potentials, respectively.
\subsection{Quench protocol}
The system is prepared in an initial state where particles are distributed uniformly in the region $\bar x\in[0,1]$. Physically, this corresponds to a zero temperature state (and thus impenetrable walls) without any drift potential ($\bar{v}_d=0$). At time $\bar t=0$, both the temperature and $\bar{v}_d$ are quenched to finite, non-zero values. Representing temperature of the system before and after the quench by $T_I$ and $T_F$, respectively, we thus have $T_I=0$ and $T_F\neq 0$. We perform numerical integration of Eq. (\ref{onel}) for $\bar{t}>0$ in order to solve the time evolution of the system following the quenches. After the quench, the particles gain kinetic energy and can penetrate into the walls; this gives rise to a mechanical force on the walls via $V_{\textrm{wall}}$.


\subsection{Coarse-grained description}
Coarse-graining of the Langevin Eq.~(\ref{onel}) via the (fluctuation-averaged) particle density, which in dimensionless form reads 
\al{
\bar{\rho}(\bar{x},\bar{t})=L \sum_{i=1}^N \delta[\bar{x}-\bar{x}_i(\bar{t})],
\label{eq:dens}
}
naturally leads to the the Smoluchowski equation \cite{kreuzer},
\begin{equation}\label{1}
\partial_{\bar{t}} \bar{\rho}(\bar{x},\bar{t}) = \partial_{\bar{x}}^2 \bar{\rho}(\bar{x},\bar{t}) - 2 \bar{v}_d \partial_{\bar{x}}  \bar{\rho}(\bar{x},\bar{t})\equiv - \partial_{\bar{x}} \bar{j}(\bar{x},\bar{t})
\end{equation}
for the density $\bar{\rho}$ and current $\bar{j}$ between the two walls ($0<{\bar x}<1$). Without drift ($\bar{v}_d=0$), \Eref{1} is simply the diffusion equation [the diffusion coefficient $D$ enters via $\bar t$; see \Eref{one}]. In the coarse-grained view, the characteristic length-scale of the wall potentials, set by $\lambda$, is taken to be much smaller than $L$, so that the walls simply impose no-flux boundary conditions on $\bar\rho$.

\subsection{Quench-induced boundary layers}
At large scales and with $\bar{v}_d=0$, the temperature quench gives rise to an initial adsorption/desorption of particles at the walls, while the density far from the walls (where the potential is zero) remains unchanged. The quench thus effectively changes the volume that the confined particles can explore by modification of the boundary layer. After the quench, these boundary layers (or depletions) diffuse into the ``bulk'' between the plates, until a homogeneous distribution is reached once more. For $\bar{v}_d=0$, the thickness of the boundary layers has been quantified by considering the initial (homogeneous) distribution and the  steady-state post-quench distribution in terms of Boltzmann weights of the wall potentials \cite{rohwer18}. In the present system, with $\bar{v}_d\neq0$, we follow a similar approach to find the initial density ${\bar \rho}(\bar{x},\bar{t}=0)$ given the boundary conditions in Eq. (\ref{1}). Before the quenches, the system is in equilibrium at a given temperature $T_I$, and there is no external potential between the walls ($\bar{v}_d=0$ for $\bar{t}<0$). If $\bar{v}_d\lesssim 1$ (weak driving compared to the post-quench thermal energy), one may assume that only the distribution of particles very close to the walls changes at very short times, while away from the boundaries the density remains unchanged. In this way, boundary layers with finite widths $\epsilon_{l,r}$ form at ${\bar x}=0,1$, representing regions with excess densities at very short times after the quenches.  For very narrow boundary layers with $\epsilon_{l,r}\ll L$, we can ignore their contributions to the boundary conditions of $\bar\rho$. Thus, on coarse-grained scales, the density obeys no-flux boundary conditions at the walls,
\begin{equation}\label{bou}
\bar{j}(\bar{x}=0/1,\bar{t})=
\bigg[ -\partial_{\bar{x}} \bar{\rho}(\bar{x},\bar{t})+ 2 \bar{v}_d \bar{\rho}(\bar{x},\bar{t})\bigg]_{\bar{x}=0,1}=0
\end{equation}
at the walls. On the same grounds, the initial coarse-grained density is uniform between the walls, excluding the boundary layers of thickness $\epsilon_{l,r}$ at the left and right sides,
\begin{equation}\label{hgh}
\bar{\rho}(\bar{x},\bar{t}=0) = \begin{cases}
  \bar{\rho}_0, & \bar{\epsilon}_r<\bar{x}<1-\bar{\epsilon}_l, \\
  0, & 0<\bar{x}<\bar{\epsilon}_r \;\textrm{or}\;1-\bar{\epsilon}_l<\bar{x}<1,
\end{cases}
\end{equation}
with $\bar{\epsilon}=\epsilon/L$. These boundary thicknesses will be estimated below. \Eref{hgh} is the pre-quench steady-state of the system (for $\bar{t}=0^-$) and fulfils the boundary condition (\ref{bou}). Immediately after the quench, the boundary layers are formed as particles start to fill those regions, i.e., the quench effectively alters the volume of the system. In Ref. \cite{rohwer18} (where $\bar{v}_d=0$), the quench was captured in terms of delta-function-like ad/desorption layers at the walls, which set the initial conditions of the boundary layer. For $\bar{v}_d\neq0$, such distributions cannot satisfy the boundary conditions in \Eref{bou}. Thus \Eref{hgh} provides a mathematically consistent alternative for the initial density, and enables us to obtain analytical results that are in excellent agreement with the simulations even at very short times.

To calculate the thicknesses of boundary layers due to the quench, we compute the pre- (${\cal R}_{I}$) and post-quench (${\cal R}_{F}$) penetration depth of the particles inside the left wall,\footnote{Specifically, ${\cal R}=\int_0^\infty dx \rho(x)/\rho(x=0)$ for a given density $\rho(x)$.}
\begin{eqnarray}\label{10}
\bar{\cal R}_{I/F}=\frac{{\cal R}_{I/F}}L=\int_{-\infty}^{0} d\bar{x} \: e^{-\frac{V_{\rm wall}(\bar{x})}{k_B T_{I/F}}}.  
\end{eqnarray}
Assuming identical wall potentials on the left and right, an analogous relation to Eq.~(\ref{10}) is obtained for  the right wall by changing the integration interval to $(1,\infty)$. Eq.~(\ref{10}) is obtained by assuming Boltzmann-type distributions long before and long after the quench, $\bar{\rho}_{I/F}(\bar{x}) \propto \exp[{-V_{\rm ex}(\bar{x})/k_B T_{I/F}}]$, and enforcing a conserved integrated density. This allows us to find the (rescaled) change in the thickness of the boundary layers induced by the quench ($\bar{\epsilon}_l = \bar{\epsilon}_r=\bar{\epsilon}$),
\begin{eqnarray}\label{ghr}
\bar{\epsilon}=\bar{\cal R}_{F}-\bar{\cal R}_{I} =\frac{\sqrt{\pi k_B T_F}-\sqrt{\pi k_B T_I}}{L\sqrt{2 \lambda }}.
\end{eqnarray}
For the quenches with $T_I=0$, the boundary layer thickness and the strength of the wall potential in \Eref{onel} are directly related by 
$\bar{\epsilon}^2=\pi/(2\bar{\lambda})$. 

\subsection{Post-quench density dynamics}
We now compute the post-quench density for $\bar t>0$ subject to the no-flux boundary conditions in \Eref{bou}, and subsequently find the pressure and forces on the walls. Since the boundary conditions are asymmetric (of ``Robin'' type), the method of images used in Ref.~\cite{rohwer18} cannot be employed here. Instead, we use separation of variables, assuming a solution of the form $\bar{\rho}(\bar{x},\bar{t})=X(\bar{x}) T(\bar{t})$. Insertion into the diffusion-drift Eq.~(\ref{1}) yields
\begin{equation}\label{2e}
\frac{\partial_{\bar{t}} T(\bar{t})}{T(\bar{t})} = \frac{1}{X(\bar{x})} \partial_{\bar{x}}^2 X(\bar{x}) - \frac{2 \bar{v}_d}{X(\bar{x})} \partial_{\bar{x}}  X(\bar{x})=-\alpha.
\end{equation}
Via the boundary conditions we obtain complete sets of spatial and temporal eigenfunctions,
\begin{eqnarray}
&&X_n(\bar{x})=e^{\bar{v}_d \bar{x}} \bigg(n \pi \cos \beta_n \bar{x} +\bar{v}_d \sin\beta_n \bar{x} \bigg),\nonumber \\
&&T_n(\bar{t})=e^{-\alpha_n \bar{t}}, \:\:\:\:\:\:\:\beta_n=(\alpha_n-\bar{v}_d^2)^{\frac{1}{2}}= n \pi,\nonumber
\end{eqnarray}
from which a general solution of \Eref{1} follows:
\begin{eqnarray}
&&\bar{\rho}(\bar{x},\bar{t})= a^{\prime} e^{2 \bar{v}_d \: \bar{x}}+\sum_{n=0}^{\infty} a_n X_n(\bar{x}) T_n(\bar{t}).
\label{4g}
\end{eqnarray}
The initial condition in Eq. (\ref{hgh}) and the orthogonality condition 
$2\int_0^1 d \bar{x}\,  \exp({-2 \bar{v}_d \bar{x}})\,  X_n(\bar{x})\,X_m(\bar{x})/(m^2 \pi^2 +\bar{v}_d^2)=\delta_{m n}$
for the spatial solutions $X_n(\bar{x})$ lead to
\begin{eqnarray}\label{momentsdynamics} 
a^{\prime}&=&\frac{\bar{v}_d e^{-\bar{v}_d} {\bar\rho}_0(1-2\bar{\epsilon})}{\sinh{\bar{v}_d}},\\
a_n&=& \frac{2 {\bar\rho_0} e^{-\bar{v}_d \bar{\epsilon}}}{(\bar{v}_d^2+n^2 \pi^2)^2}\bigg[ (1+W_n) (\bar{v}_d^2-n^2 \pi^2)\sin({ n \pi \bar{\epsilon}})\nonumber\\
&&-2 \bar{v}_d n \pi (W_n-1) \cos({ n \pi \bar{\epsilon}})\bigg],\quad
\end{eqnarray}
with $
W_n = (-1)^{n} \exp[{\bar{v}_d(2 \bar{\epsilon}-1)}]$.

\subsection{Pressure}\label{pressure}
For an ideal gas of Brownian particles, pressure on a confining surface follows from the contact density~\cite{solon_pressure_2015PRL,rohwer18}, $P = k_B T_F \rho$, i.e.,
\begin{eqnarray}\label{tir}
\bar{P}_{l/r}(\bar{x}=0/1,\bar{t})
=\frac{P_{l/r} }{({\rho}_0 k_B T_F)}
=\frac{\bar{\rho}(\bar{x}=0/1,\bar{t})}{\bar{\rho_0}}, 
\end{eqnarray}
where ${\rho_0}$ is the pre-quench density between the walls. In a 1D system, and thus in our simulations, ${\rho_0}\equiv \bar{\rho_0}/L$. Thus Eq. (\ref{4g}) facilitates an analytical formula for the pressure exerted on the left/right walls ($\bar{x}=0/1$) in the coarse-grained system.

On the other hand, in explicit simulations of the Langevin Eq.~(\ref{onel}), pressure on the walls can be obtained in terms of mechanical forces. For the left wall, 
$P_l = \int_{-\infty}^0 dx\, \rho(x)V'_{\rm wall}(x)$, 
and analogously for the right wall. Using \Eref{eq:dens}, one finds
\begin{eqnarray}
\bar{P}_{l}  = \frac{-\bar{\lambda}}{\bar{\rho_0}}\sum_{i=1}^N  \bar{x}_i \Theta(-\bar{x}_i), \;\,\,\,
\bar{P}_r = \frac{\bar{\lambda}}{\bar{\rho_0}}\sum_{i=1}^N   (\bar{x}_i\!-\!1) \Theta(\bar{x}_i\!-\!1).
\end{eqnarray}
\begin{figure}[t!]
\centering
\includegraphics[width=0.98\linewidth]{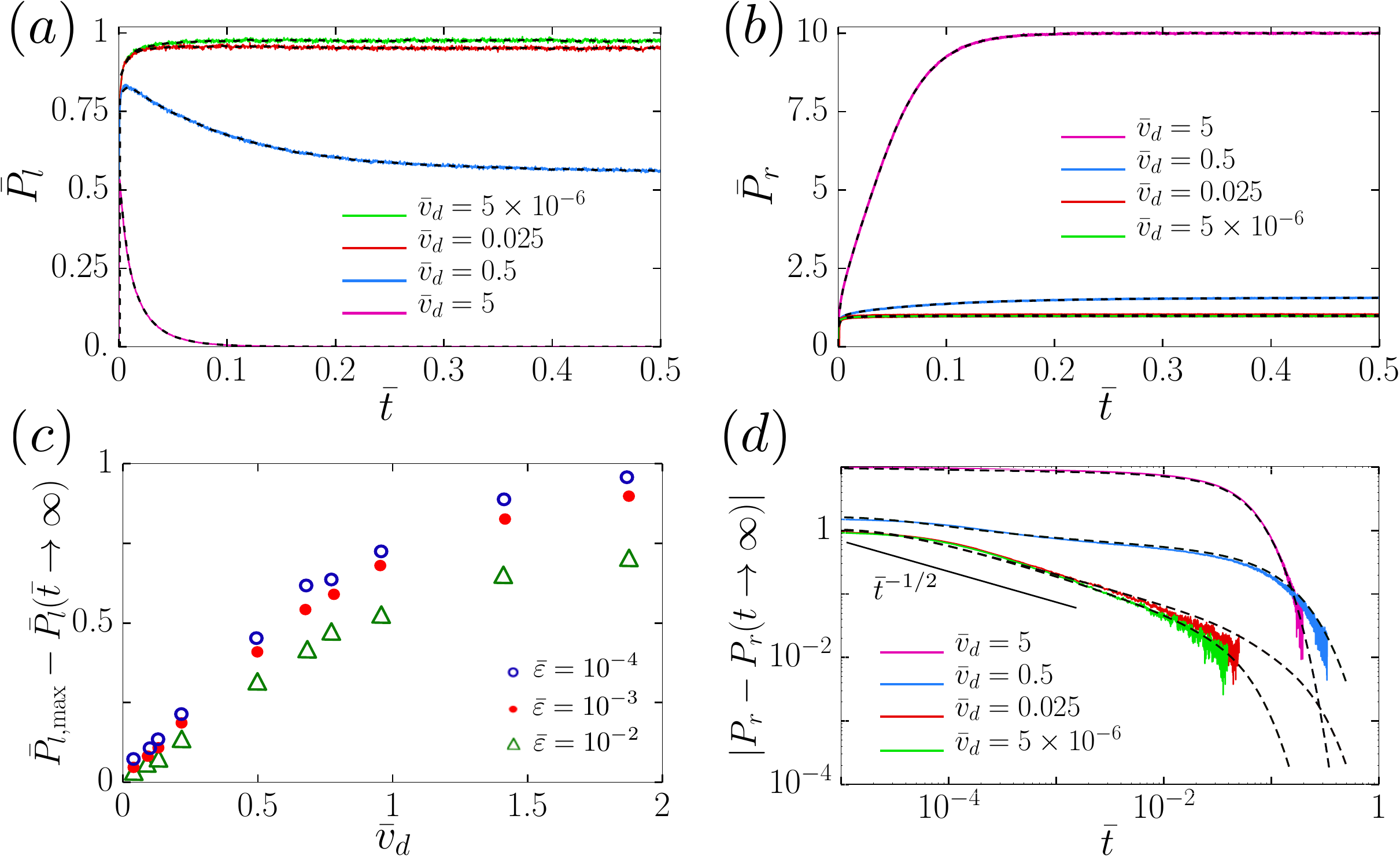}
\caption{(Color online) Pressure on the right (a) and left (b) wall, as a function of $\bar t = D t/L^2$, for different values of the effective drift $\bar{v}_d$. Solid (colored) lines are simulation results from Langevin dynamics, while dashed (black) lines represent the analytical description [see \Eref{tir}].
(c) Difference between the peak and the steady-state pressures on the left wall as a function of $\bar{v}_d$ for different $\bar{\epsilon}$.
(d) Approach of the pressure on the right wall to its stationary value. In panels (a), (b), and (d) the effective boundary layer thickness is $\bar{\epsilon}=0.01$.
}
\label{fig2}
\end{figure}

\section{Results}
Following the quench at $t=0$, the driving force explicitly breaks the left/right symmetry of the system. Consequently the two walls experience different dynamic forces and pressures, as shown in Fig.~\ref{fig2}. For pressure, we present both analytical coarse-grained results as well as numerical simulations for $N=10^7$ particles, which are in excellent quantitative agreement with each other. 
Since the drift velocity points towards the right wall ($\bar{v}_d>0$), the pressure on the right wall increases monotonically as more particles penetrate into the wall, until a steady-state value is reached. In contrast, time evolution of the pressure on the left wall is non-monotonic, exhibiting an ``overshooting'' behavior. Shortly after the quench, particles penetrate into the wall so that the pressure increases. At later times, the pressure decreases as drift removes particles from the left wall, until the long-time stationary state is reached. As shown in Figs.~\ref{fig2}(a) and \ref{fig2}(b), an increase of the external field strength (i.e., of the drift velocity $\bar{v}_d$) leads to an enhancement not only of the difference between pressures on the two walls, but also of the relative amplitude of the left pressure overshoot. For very small values of $\bar{v}_d$, the evolution of density on the right and left walls becomes equivalent as symmetry is restored. 
\par
Fig.~\ref{fig2}(c) addresses the overshooting behavior, showing the difference between the maximum value and the steady-state value of the non-equilibrium pressure at the left wall, in dependence on $\bar{v}_d$, for various $\bar{\epsilon}$.\footnote{\,$\bar\epsilon$ captures the magnitude of the temperature quench via \Eref{ghr}.} 
The overshooting becomes more pronounced with increased drift: ${\bar P}_{l,{\rm max}}-{\bar P}_l({\bar t}\to\infty)\to 1$ when the boundary layer thickness becomes very small ($\bar{\varepsilon}=10^{-4}$). For larger values of $\bar\epsilon$ (i.e., less steep wall potentials / larger temperature quenches), the overshooting is slightly reduced, because the particles can penetrate deeper into the wall before being removed by drift. 
While the time evolution of pressure on the right wall is monotonic in the presence of the external field, its approach to the steady regime depends strongly on $\bar{v}_d$, as shown in Fig.~\ref{fig2}(d).
At intermediate times ($10^{-4}\lesssim\bar{t}\lesssim10^{-2}$), the pressure difference $\bar{P}_r(\bar{t})-\bar{P}_r(\bar{t}\to\infty)$ follows a power law decay $\sim\bar{t}^{-\zeta}$, where the exponent $\zeta$ changes from $1/2$ to a vanishingly small value upon varying the external field strength from a very small $\bar{v}_d = 10^{-6}$ to $\bar{v}_d=1$. Because of the overshooting effect in ${\bar P}_{l}({\bar t})$, no power law behavior is observed for the left wall. At late times, the pressure on both walls decays exponentially with time towards the stationary regime. 

\begin{figure}[t!]
\centering
\includegraphics[width=0.99\linewidth]{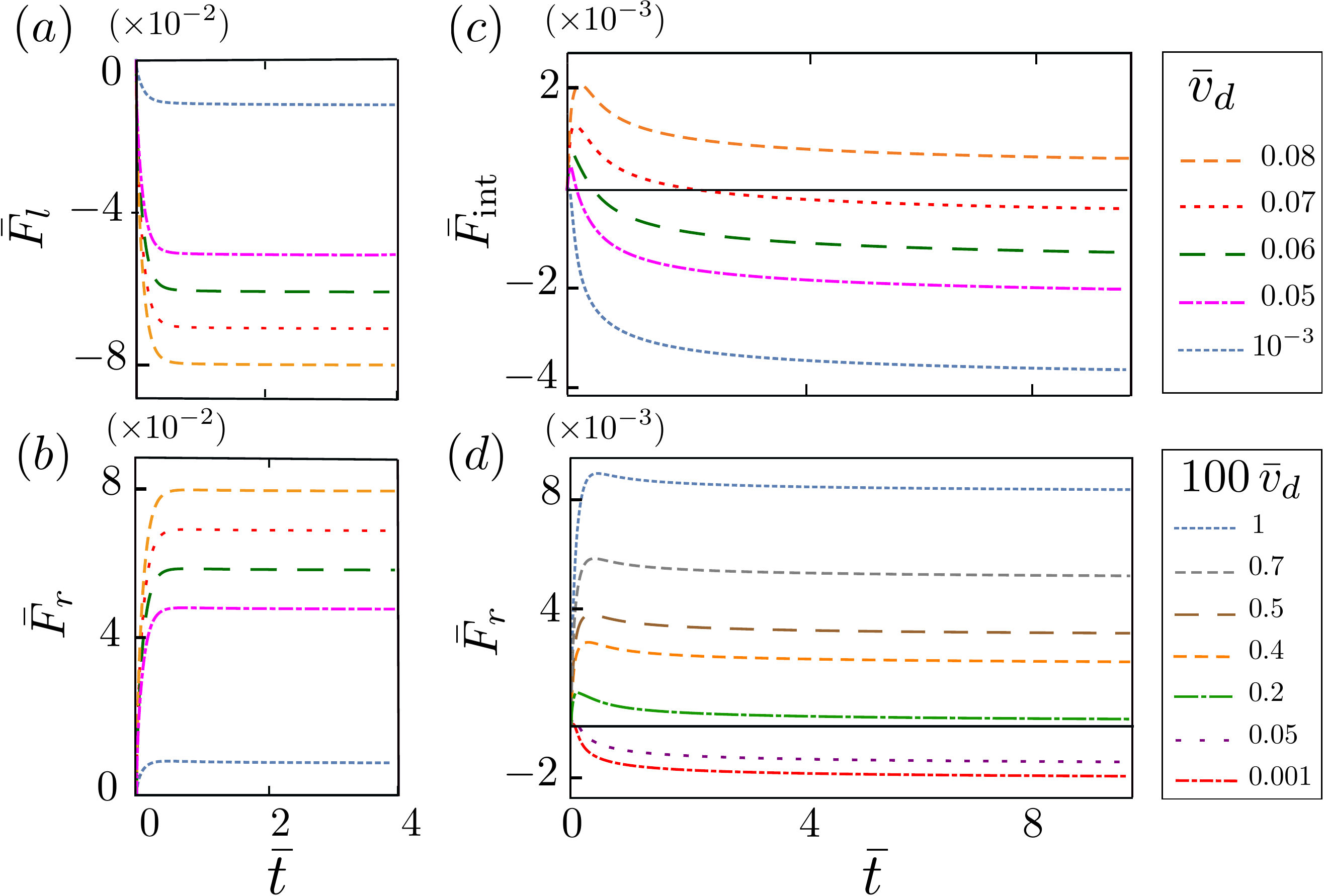}
\caption{(Color online)
Analytical results: time evolution of the force on the left (a) and the right (b,d) walls, and the effective interaction force $\bar F_{\textrm{int}}=\bar{F}_{r}+\bar{F}_{ l}$ (c), following the quenches, for different values of $\bar{v}_d$, with $\bar t = D t/L^2$.
Positive values of $\bar{F}_{r/ l}$ imply that the force points to the outside of the plates, while positive values of $F_{\textrm{int}}$ indicate an effective repulsion of the plates. For the interval of the external field strength $\bar{v}_d$ shown in panel (c), the effective interaction force changes sign in time. Panel (d) shows the force on the right wall for very small values of $\bar{v}_d$. All results here are obtained for $\bar{\epsilon}=10^{-3}$. The legend of panel (c) also applies to (a) and (b).}
\label{fig3}
\end{figure}

We now study the net forces exerted on the plates when they are immersed in an infinite Brownian fluid. This requires knowledge of the density evolution  outside the plates following the temperature quench. Since the drift potential is set to zero outside, the initial condition on the densities external to the plates, $\bar{\rho}_{\rm out}(\bar{x},\bar{t}=0)$, can be obtained from semi-infinite versions of \Eref{hgh} for the left and right outside regions: one has a constant density $\bar\rho_0$ away from the walls, and zero density inside the adsorption layers of thickness $\bar\epsilon$ on each wall. Together with Eq.~(\ref{4g}), 
this yields the corresponding expansion coefficients. The net force on each wall is now obtained as the \textit{difference of inside and outside pressures},
\begin{eqnarray}\label{ttir}
\bar{F}_{l/r}(\bar{t})= \frac{1}{\bar{\rho}_0}\left[ \bar{\rho}(\bar{x}=0/1,\bar{t})-\bar{\rho}_{\rm out}(\bar{x}=0/1,\bar{t}) \right] ,
\end{eqnarray}
where we have used the definition $\bar{F}= A F /(\rho_0 k_B T_F)$ for the dimensionless force.\footnote{Here the area $A$ of the walls is $d-1$ dimensional, i.e., in a 1D system, pressure and force have the same dimension.} We define an \textit{effective} interaction force between the plates as $\bar{F}_{\rm int}(\bar{t})= \bar{F}_{l}(\bar{t})+\bar{F}_{r}(\bar{t})$ which becomes positive (negative) if the plates experience an effective repulsion (attraction). 
\begin{figure}[t!]
\centering
\includegraphics[width=0.85\linewidth]{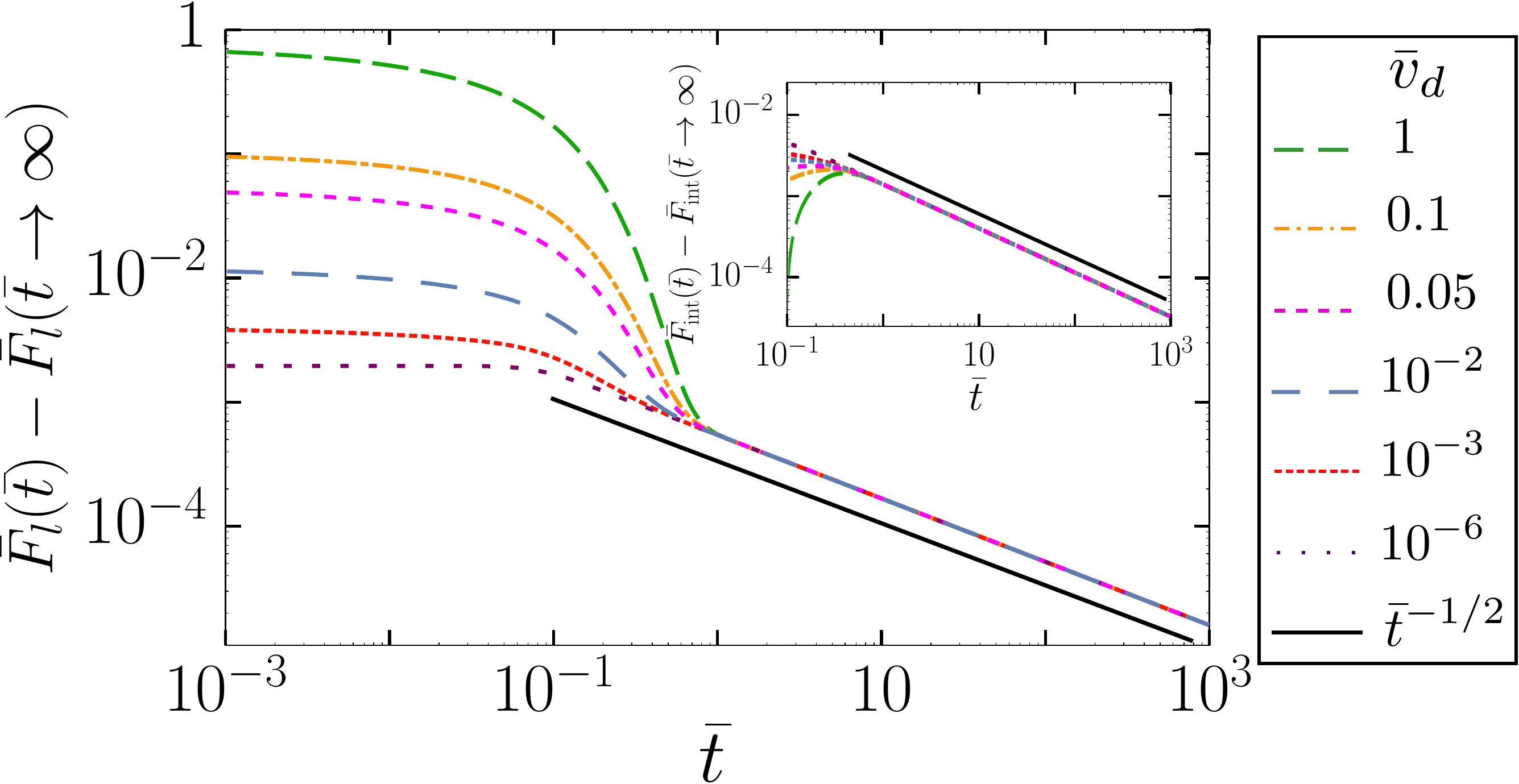}
\caption{(Color online) Analytical results: approach of the force $\bar F_{l}$ on the left wall towards its steady-state value following the quench, for different effective drift strengths $\bar{v}_d$, with $\bar\epsilon = 10^{-3}$ and $\bar t = D t/L^2$. Inset: a similar plot for the effective interaction force $\bar F_{\textrm{int}}$ for times after the overshoot has occurred. 
}
\label{fig4}
\end{figure}
\par
The post-quench time evolution of the forces is shown in Fig.~\ref{fig3} for various $\bar{v}_d$.
For sufficiently large values of $\bar{v}_d$, the post-quench force on the left wall points towards the inside, while the force on the right wall is towards the outside.  This occurs since drift transports particles from the left wall to the right wall. 
Fig.~\ref{fig3}(c) shows that the effective interaction force between the plates can change its sign at short times following the quenches, since the left and right forces evolve asymmetrically. In fact, while the long-time interaction force is always negative (attractive), at short times after the quenches, the net force can be positive (repulsive). This behavior can be ascribed to the interplay of the quench and the external field (drift), and the fact that dynamics inside and away from the walls have different characteristic time scales. The crossover time from repulsive to attractive interaction forces strongly depends on the strength of external field:  for larger values of $\bar{v}_d$, the repulsion is retained until much later times. Concerning the net forces on the individual plates, Fig.~\ref{fig3}(d) reveals that by decreasing the external field intensity to small values $\bar{v}_d\lesssim 10^{-2}\,-\,10^{-5}$, an overshooting feature appears in the force on the right plate (in direct contrast to the case of pressure, where the overshooting was observed for left wall pressure). In addition, for very small values of $\bar{v}_d$, the steady-state force is towards the outside of the right plate ($\bar{F}_r<0$), which is consistent with the result obtained in Ref. \cite{rohwer18}: in the absence of drift, repulsive net forces occur at late times.
Figure \ref{fig4} shows how the force on the left plate and the effective interaction force approach their steady states. 
The force on the left wall does not exhibit overshooting. Both $\bar F_l$ and $\bar F_{int}$ have two dynamic regimes: At short times these forces decay towards their steady state exponentially quickly, but after diffusion across the plate separation has occurred (${\bar t}\sim1$), decay crosses over to $\bar{t}^{-1/2}$. 
This power-law decay at late times has been explicitly shown for drift-free ($\bar{v}_d=0$) post-quench dynamics \cite{rohwer18}. Thus we infer that drift-related contributions are relevant at short times, before the $\sim\bar{t}^{-1/2}$ scaling  appears.

\section{Conclusions}
We have studied dynamics of the density of an ideal Brownian gas between two plates, as well as the associated non-equilibrium forces, following simultaneous quenches of the temperature and an external driving field. Our analytical coarse-grained results and numeric simulations show excellent agreement. 
At intermediate post-quench times, the presence of drift leads to strong deviations of the dynamics from the purely diffusive case. For drift towards the right, the pressure on the left wall exhibits a non-monotonic time evolution, while on the right wall we see a monotonic increase, approaching the steady pressure as $\sim\bar{t}^{-\zeta}$. When the drift strength $\bar v_0$ is increased from zero, $\zeta$ decreases from $1/2$ to $0$ as drift begins to dominate over diffusion. Further, for two walls embedded inside the (bulk) gas, we studied the net forces on each wall, as well as an effective interaction force between the walls. At late times, the effective interaction force $\bar F_{int}$ between the plates is always attractive, and decays as $\sim t^{-1/2}$ to its steady value, indicating diffusion-dominated dynamics. At early times, however, the drift-induced overshoot on the left wall renders $\bar F_{int}$ repulsive for certain values $\bar v_d$.

We expect the quenches and external fields described here to be realizable in a variety of physical systems, including colloidal suspensions, active matter and electrically charged fluids.

\acknowledgments
We are grateful to P. Nowakowski for valuable discussions. Further, we thank S. Dietrich for funding a research visit of MRN at the MPI-IS in Stuttgart, during which this work was completed, as well as for financial support of CMR.

\bibliographystyle{eplbib}
\bibliography{knmr.bib}

\end{document}